\documentclass[11pt]{article}
\usepackage{amsfonts}
\usepackage{amssymb}
\def\bsh{\backslash}

\def\adt{\dot \alpha}

\newfont{\bbbold}{msbm10 scaled \magstep1}

\def\bbC{\mbox{\bbbold C}}

\def\bbP{\mbox{\bbbold P}}

\def\cG{{\cal G}}

\def\cO{{\cal O}}

\newfont{\goth}{eufm10 scaled \magstep1}

\def\a{\alpha}

\def\d{\delta}\def\D{\Delta}

\def\l{\lambda}

\def\p{\pi}

\def\be{\begin{equation}}\def\ee{\end{equation}}
\def\bea{\begin{eqnarray}}\def\eea{\end{eqnarray}}
\def\ba{\begin{array}}\def\ea{\end{array}}

\def\str{\rm str}

\def\bd{\begin{document}}
\def\ed{\end{document}}
\def\bea{\begin{eqnarray}}
\def\eea{\end{eqnarray}}
\def\ba{\begin{array}}
\def\ea{\end{array}}
\def\ft#1#2{{\textstyle{{\scriptstyle #1}\over {\scriptstyle #2}}}}
\def\fft#1#2{{#1 \over #2}}
\newcommand{\eq}[1]{(\ref{#1})}
\def\eqs#1#2{(\ref{#1}-\ref{#2})}
\def\det{{\rm det\,}}
\def\tr{{\rm tr}}
\newcommand{\ho}[1]{$\, ^{#1}$}
\newcommand{\hoch}[1]{$\, ^{#1}$}
\newcommand{\tamphys}{\it\small Center for Theoretical Physics,
Texas A\&M University, College Station, TX 77843, USA}
\newcommand{\newton}{\it\small Isaac Newton Institute for Mathematical
Sciences, Cambridge, UK}
\newcommand{\kings}{\it\small Department of Mathematics, King's College,
London, UK}
\newcommand{\lapp}{\it\small LAPP, Annecy, France}
\begin{document}
 \thispagestyle{empty}
 
\hfill{IASSNS-HEP-99-93}

 \hfill{KCL-MTH-99-44}

 \hfill{LAPTH-754/99}

 \hfill{\today}
\vspace{20pt}

 \begin{center}
{\Large{\bf Extremal correlators in four-dimensional SCFT}}
\vspace{30pt}\\ {\large B. Eden\hoch{a,b}, P.S. Howe\hoch{a}, C. 
Schubert\hoch{b,c}, E. Sokatchev\hoch{b} and P.C. West\hoch{a}} 
\vspace{15pt} 
\begin{itemize}
 \item [$^a$] \kings
 \item [$^b$] {\it\small Laboratoire d'Annecy-le-Vieux de Physique
 Th{\'e}orique\footnote{UMR 5108 associ{\'e}e {\`a}
 l'Universit{\'e} de Savoie} LAPTH, Chemin de Bellevue, B.P. 110,
F-74941 Annecy-le-Vieux, France}
\item [$^c$] {\it\small School of Natural Sciences, 
Institute for Advanced Study, Olden Lane, Princeton, NJ 08540, US}  
 \end{itemize}
\vspace{60pt}
 {\bf Abstract}
\end{center}
It is shown that certain extremal correlators in four-dimensional 
$N=2$ superconformal field theories (including $N=4$ 
super-Yang-Mills as a special case) have a free-field functional 
form. It is further argued that the coupling constant dependence 
receives no correction beyond the lowest order. These results hold 
for any finite value of $N_c$. 

{\vfill\leftline{}\vfill 
\pagebreak \setcounter{page}{1}
%%%%%%%%%%%%%%%%%%%%%%%%%%%%%%%%%%%%%%%%%%%%%%%%%%%%%%%%%%%%%%%%%%%%%%%%%%%%%%

\section{Introduction}

In an interesting recent article D'Hoker et al \cite{dfmmrext} 
have shown that certain ``extremal" correlation functions are free 
in AdS supergravity. These authors also gave some arguments for 
this to be true for the CFT correlators corresponding to those 
supergravity correlators according to the Maldacena conjecture 
\cite{maldacena}. Further one-loop and instanton checks have been 
given in \cite{biakov}. In this note we show that the 
non-renormalisation of such correlators can be understood very 
simply in non-perturbative superconformal field theory as a 
consequence of the superconformal Ward identities and the 
constrained nature of the harmonic superfields that appear in the 
theory. We further show that these simplifications hold in any 
finite $N=2$ SCFT and not just the $N=4$ SYM theory (at least for 
the correlators of $N=2$ matter multiplets). We also argue that 
the result could be extended to another class of ``subextremal" 
correlators. 

The reason that the Ward identities in superconformal theories are 
stronger than one might naively expect is not because some of the 
invariants, i.e. cross-ratios, that occur in 
non-su\-per\-sym\-met\-ric conformal field theories do not have 
supersymmetric extensions, but rather because all the basic field 
strengths and matter multiplets in four-dimensional supersymmetric 
theories are described by constrained superfields. These 
constraints, when taken together with the superconformal Ward 
identities, may lead to much stronger results than those that 
follow for non-su\-per\-sym\-met\-ric theories from conformal 
invariance alone.  For example,  the chiral nature of the matter 
superfield in $N=1$ theories can be exploited to determine the 
anomalous dimension of operators composed from this field 
\cite{n}. The same is true for the gauge field strength 
superfields in $N=1$ and $N=2$ Yang-Mills. The remaining 
supermultiplets that occur in four-dimensional theories with rigid 
extended supersymmetry are the $N=2$ matter multiplet and the 
$N=4$ Yang-Mills multiplet. The former  has an off-shell   
harmonic superspace  formulation \cite{hh} which can be used to 
give a complete description of the quantum theory including its 
Feynman rules and superconformal properties. While only an 
on-shell harmonic superspace formulation \cite{hhh} is known for 
the latter it can still be used to investigate the properties of 
the correlation functions under superconformal transformations.  
Both of these multiplets  are described by superfields which 
satisfy simple constraints:  the superfields obey the condition of 
Grassmann (G-)analyticity (an analog of chirality in $N=1$ 
theories) and are harmonic (H-)analytic (a dynamical property of 
$N=2$ matter \cite{hh} which can also be extended to $N=4$ SYM 
\cite{hhh}). These constraints mean that such superfields depend 
on only half of the number of Grassmann odd coordinates in the 
full superspace, as in the case of chiral superfields, and that 
they depend analytically  on the coordinates of the internal coset 
space which is adjoined to super Minkowski space to form harmonic 
superspace. We shall refer to such superfields as analytic 
superfields in the following. 

In references \cite{hw1,hw2} a systematic study of the 
superconformal Ward identities and their consequences for Greens 
functions of $N=2$ and $N=4$ analytic operators was initiated. The 
$N=2$ operators are gauge-invariant products of the hypermultiplet 
$q$ and the $N=4$ operators are gauge-invariant products of the 
$N=4$ field strength $W$. (We shall use the notation $\cO_p$ to 
denote operators of the form $\tr(q^p)$ or $\tr (W^p)$.) This 
approach has led to a number of interesting results 
\cite{hw1,hw2}, in particular: the $N=4$ SYM field strength is a 
covariantly analytic scalar superfield $W$ from which the 
aforementioned set of analytic gauge-invariant operators can be 
built; the two- and three-point Greens functions of these 
operators were determined up to constants \cite{hw1,hsw} (see also 
\cite{ken2}); the set of all non-nilpotent analytic superconformal 
invariants was given \cite{hw2}. It was also conjectured  that the 
superconformal Ward identities and the constrained nature of the 
harmonic superfields might be sufficiently strong to determine 
certain correlation functions up to constants \cite{hw1,hw2}. 

It has subsequently emerged that the $N=4$ operators considered  
above play an important r\^{o}le in the Maldacena conjecture and are 
in one-to-one correspondence with the Kaluza-Klein multiplets of 
IIB supergravity on $AdS_5\times S^5$ \cite{af}. However, although 
the Ward identities, when  combined with Grassmann and harmonic 
analyticity, do put constraints on the functions of 
superinvariants that can arise for correlators with four or more 
points, these constraints do not seem to be enough 
\cite{eetal,toappear} to determine completely the $N=2$ 
correlators that contain four harmonic composite matter fields of 
charge two, contrary to  the claim for future work  made in 
\cite{hw1,hw2}. The situation is very different, however, for 
four-point correlation functions of analytic operators for which 
the sum of three of the charges (i.e. powers of $W$ or $q$ in a 
given operator) is equal to the charge on the fourth. These are 
the extremal correlators discussed in \cite{dfmmrext,biakov}. 

In this paper  we consider such extremal correlation functions in 
$N=4$ or $N=2$ SYM theories. They have the form 
$$<\cO_{p_1}\cO_{p_2}\cO_{p_3}\cO_{p_4}>
$$
where $p_4=\sum_{i=1}^3 p_i$. We shall show that the 
superconformal Ward identities together with analyticity imply 
that they can be solved for and that they can be expressed as 
products of two-point functions. This will demonstrate part of the 
conjecture of D'Hoker et al \cite{dfmmrext}, but it is a stronger 
result in the sense that it is valid for any gauge group $SU(N_c)$ 
and not only in the limit $N_c\to \infty$. Furthermore, these 
results hold also for any $N=2$ SCFT and not just $N=4$, at least 
for composite operators made out of hypermultiplets. The reason 
why these correlators are soluble, while the case $p_1=\ldots =p_4 
= 2$ considered in \cite{eetal,toappear} is not, is related to  
their different forms when expressed in terms of products of 
two-point functions multiplied by a function of superconformal 
invariants. The singularities of the latter in the internal 
coordinates must be compensated for by the zeros of the former, 
because all correlation functions have to be analytic. For the 
extremal correlators the structure of the two-point function 
prefactor allows essentially no freedom for the function of 
invariants; it must be a constant. 

The superconformal Ward identities alone do not determine the 
dependence of a correlator on the coupling constant. However, 
information on this dependence can be extracted by using the 
reduction formula, first introduced in the context of 
four-dimensional SCFT by Intriligator \cite{ken1}. This formula 
relates the derivative of an $n$-point function with respect to 
the coupling to an $(n+1)$-point function with one insertion of a 
bilinear operator and an integration over the insertion point. In 
the $N=4$ case the inserted operator is the supercurrent $T=\cO_2$ 
while in the $N=2$ case the insertion into correlators of 
G-analytic superfields is the chiral pure SYM Lagrangian. 
Consideration of this reduction formula led to the realisation 
\cite{ken1} that there must be nilpotent superinvariants since 
only such invariants can contribute in the integration over the 
insertion point. Nilpotent invariants exist for correlators with 
five or more points \cite{ehw,hssw}. Their absence for $(n\leq 
4)$-point correlators has been used to argue that two-and 
three-point correlators of analytic operators are not renormalised 
\cite{ehw}. Here we shall use the reduction formula to argue that 
the extremal four-point correlators are not renormalised as a 
consequence of the absence of nilpotent invariants with the 
required analyticity properties. We also show that the same 
argument implies the non-renormalisation of another class of 
operators, ``subextremal" for which $p_4=\sum_{i=1}^3 p_i - 2$. 
Note, however, that when using the reduction formula we make the 
(plausible) assumption that there are no undesired contributions 
from contact terms \cite{hssw,petske}. 

\section{Extremal correlators as constrained superconformal\\ covariants}

Let us begin by considering an $n$-point correlator in $N=2$ or 
$N=4$ SCFT for the operators $\cO_{p}$ discussed above. Such a 
correlator, if its leading term in an expansion in odd coordinates 
does not vanish,  can be written in the form 
\be
\cG:=<\cO_{p_1} \cO_{p_2} \ldots \cO_{p_n}> =\prod_{r<s} 
(g_{rs})^{m_{rs}} F \ee where $F$ is some function of the 
invariants, the $m_{rs}\equiv m_{sr}$ are non-negative integers 
and the $g_{rs}$ are the free two-point functions of charge one 
operators (in $N=2$ these are just the hypermultiplet 
propagators). It is  easy to see that this form for $\cG$ does 
solve the superconformal Ward identities because the propagator 
satisfies 
\be
(V_r + V_s + \D_r + \D_s) g_{rs}=0
\ee
where $V$ is the vector field generating a superconformal
transformation in analytic superspace and $\D$ is the associated
analytic density function, and because the transformation rule  for a
field of charge $p$ is
\be
\d \cO_p= V \cO_p + p\D \cO_p \, .
\ee
The $m_{rs}$ must obey the equations
\be
p_r= \sum _{s,\ r\neq s} m_{rs}
\ee
whence it follows that
\be
-p_n+\sum _{r=1}^{n-1} p_r= \sum _{r,s=1}^{n-1}m_{rs} \, . \ee In 
the extremal case $p_n=\sum _{r=1}^{n-1} p_r$ the solution to this 
equation is unique. Thus we have 
\be
\label{unique} <\cO_{p_1} \cO_{p_2} \ldots \cO_{p_n}> 
=\prod_{r=1}^{p-1} (g_{rn})^{p_r} F\quad \mbox{for}\ p_n=\sum 
_{r=1}^{n-1} p_r \, . \ee 

To be more concrete we now focus on four-point extremal 
correlators in the $N=2$ theory. The simplest example is 
$\cG:=<\cO_2(1)\cO_2(2)\cO_2(3)\cO_6(4)>$ where the $\cO$'s are 
made out of hypermultiplets. It can be written in the following 
form: 
\be
\cG= (g_{14})^2 (g_{24})^2 (g_{34})^2 F \ee where $F$ is a 
function of super-invariants. The two-point function $g_{rs}$ is 
the propagator for free hypermultiplets, 
\be
g_{rs}={\hat y_{rs}\over x_{rs}^2}
\ee
where $x_{rs}=x_r-x_s$, etc., and $\hat
y_{rs}=y_{rs}-\p_{rs}x^{-1}_{rs}\l_{rs}$. The variables $\l$ and $\p$
are the fermionic coordinates of analytic superspace carrying undotted
and dotted spinor indices respectively and $y$ is the standard local
coordinate on the coset space $U(1)\bsh SU(2)=\bbC \bbP^1$.

Let us now discuss the function $F$ depending on the 
superconformal invariants. For four points there are no nilpotent 
invariants \cite{ehw}, which means that the supersymmetric 
extension of an ordinary Minkowski or internal space invariant is 
unique, or in other words that all invariants are specified by 
their body. The two non-trivial invariants with respect to the $x$ 
conformal transformations are the cross-ratios 
\be
s={x_{14}^2 x_{23}^2\over x_{12}^2 x_{34}^2}
\qquad t={x_{13}^2 x_{24}^2\over x_{12}^2 x_{34}^2}
\ee
and the non-trivial invariant with respect to the $y$ conformal
transformations is
\be
v={y_{14}y_{23}\over y_{12}y_{34}}\, . \ee The quantity $w ={ 
y_{13} y_{24}\over y_{12} y_{34}}$  is conformally invariant, but 
can be written in terms of $v$, namely $w=1+v$. Thus for four 
points in analytic superspace, there are three independent 
invariants. Any basis set whose body can be solved for the 
cross-ratios $s,t,v$ may be chosen, for instance \cite{hw2} 
\be
S={s\det X_{14}s\det X_{23}\over
s\det X_{12}s\det X_{34}},\
T={s\det X_{13}s\det X_{24}\over
s\det X_{12}s\det X_{34}}
\ee
and
\be
U=\str (X^{-1}_{34}X_{41}X^{-1}_{12}X_{23}) \ee where $X$ is a 
$(2|N)$ by $(2|N)$ matrix coordinatising a patch on analytic 
superspace and $X_{rs}$ denotes a coordinate difference. 
\par
The superconformal invariants $S,T$ can be written as \be S= 
{s\over \hat v},\  T = {t \over \hat w} \ee where \be \hat v={\hat 
y_{23}\hat y_{14}\over\hat y_{12}\hat y_{34}},\ \ \hat w={\hat 
y_{13}\hat y_{24}\over\hat y_{12}\hat y_{34}} \, . \ee It is clear 
that both of these objects and also $U$ contain singularities in 
$y$. At the same time, a correlation function of gauge-invariant 
composite operators made out of hypermultiplets must be analytic 
in $y$. This implies that the function $F$ in equation 
(\ref{unique}) must depend on the superinvariants in such a way 
that the singularities in $y$ that appear in $F$ are cancelled 
by the zeros in $y$ contained in the propagators. Furthermore, 
this must be true to all orders in $\lambda\pi$. To examine this 
issue we will need to know the dependence of the superinvariants 
on $y$ and $\lambda,\pi$. It turns out that the singularities 
cannot be cancelled in the extremal case, so that the function $F$ 
can only be a constant. 
\par
We write the nilpotent parts of  $\hat w$ and the superconformal
invariant $U$ as $\Delta w$ and $\Delta U$ respectively, namely
\be
\hat w = 1+\hat v + \delta w,\qquad
U=1-t+s+\hat v+\Delta U \, .
\ee
\par
According to the argument above we can, by taking appropriate 
functions of $S,T,U$, construct three superconformal invariants 
whose leading terms are given by $s, t$ and ${v\over s}$ . These 
superconformal invariants are 
\be
S^\prime=SV,\qquad T^\prime=T(1+V),\qquad z={1\over S} \ee where 
$V={T+U-1\over 1+S-T}$. To first order in $\Delta w$ and $\Delta 
U$ we find \bea S^\prime&=&s + \Delta_s = s-{t\over R}{\Delta 
w\over(1+sz)} +{  \Delta U\over R}+\ldots,\\ 
 T^\prime&=&t + \Delta_t
=t+(1+z)-{t\over R}{\Delta w\over(1+sz)}+{zt\over(1+sz)}{  \Delta
U\over R} + \ldots, \\ z&=&{\hat v\over s} \eea with 
$R=(1+z)-{zt\over 1+sz}$. 
\par
Finally, we are in a position to show that $F$ is a constant as a 
consequence of the analyticity of the extremal correlators in 
their dependence on $y$. Taylor-expanding to first order in 
$\Delta U$ and $\Delta w$ we find that 
\be
F(s+  \Delta_s,t+ \Delta_t,z) =F(s,t,z)+{  \Delta U\over R} D 
F-{t\Delta w\over  R(1+sz)}\hat D F+\ldots \ee where 
\be
D:={\partial\over\partial s} +{zt\over1+sz}{\partial\over\partial 
t}\;,\qquad \hat D:={\partial\over\partial 
s}+(1+z){\partial\over\partial t}\;. \ee 
\par
Now, the correlation function must be analytic, but the variable 
$z$ contains a singularity in $y_{12}y_{34}$. Strictly speaking, 
we consider the singularities in  $\hat y_{12}\hat y_{34}$. It is 
a special feature of the extremal case that the factors of $\hat 
y$ in the two-point functions that multiply $F$ can never cancel 
such a singularity  and hence $F$ does not depend on $z$. We now 
consider the terms in $F$ which are first order in $\lambda\pi$, 
that is, in $\Delta U$ and $\Delta w$. Since these terms have 
independent spinor structures and go like $v$ and we are taking $v 
\to 0$, we must conclude that 
\be
D F = 0,
\ee
and
\be
\hat D F= 0 \, . \ee By considering the coefficient of each power 
of $z$ in the first equation, it is straightforward to show that 
$F$ is actually a constant. Clearly, the second equation is then 
automatically satisfied. 
\par
The same conclusion can be reached using $N=4$ harmonic superspace 
for a similar correlator with three charge two operators (i.e. 
supercurrents) and one charge six operator. In this case one has 
the same form for $\cG$ but the two-point function is now 
different. It takes the form 
\be
g_{rs}={\hat y_{rs}^2\over x_{rs}^2} \ee where $\hat 
y_{rs}=y_{rs}-\p_{rs}x^{-1}_{rs}\l_{rs}$ as  before, the only 
difference being that the odd coordinates and $y$ now carry 
internal indices as well. The coordinates of $N=4$ analytic 
superspace are $(x^{\a\adt},\l^{\a a'},\p^{a\adt},y^{a a'})$ where 
each index can take on two values and the internal indices are 
treated in the same way as the spacetime spinor indices. In the 
$N=4$ case there are four independent super-invariants at four 
points corresponding to the fact that there are two independent 
spacetime cross-ratios and two independent internal space 
cross-ratios each of which has a unique extension to a 
super-invariant. Again, all of these are singular in such a way 
that the singularities cannot be cancelled by the prefactor $\Pi 
g^m$ in the extremal correlator, and hence we conclude that the 
only function of super-invariants that one can have is in fact a 
constant. 

\section{Four-point correlators and the reduction formula}

We now consider the question of non-renormalisation of the 
coefficient multiplying the propagators in the expression for the 
four-point correlators discussed above. This can be done either in 
the $N=4$ formalism or in $N=2$. In this note we shall give the 
$N=2$ version since it is applicable to an arbitrary $N=2$ 
superconformal theory. In this sense the results obtained are 
stronger than the $N=4$ ones, although they only apply to the  
hypermultiplet sector of the theory. Furthermore, in this section 
we shall use the manifestly $SU(2)$-covariant harmonic superspace 
formalism of reference \cite{hh}. 

The four-point correlators we discuss involve gauge-invariant 
composite operators $\cO_{p_r} = \mbox{tr}(q^{p_r})$ made out of 
$p_r$ hypermultiplets: 
\begin{equation}\label{1}
  \langle \cO_{p_1}(1)\ldots \cO_{p_4}(4)\rangle \equiv \langle 
p_1p_2p_3p_4\rangle\;. 
\end{equation}
The requirement of gauge invariance puts the natural restriction 
$p_r\geq 2$ on the allowed values of the charges. As stated 
earlier, such correlators should have two fundamental properties: 
superconformal covariance (follows from the finiteness of the 
theory) and harmonic (H-)analyticity. The latter is a dynamical 
property which can be explained as follows.

In the $SU(2)$-covariant harmonic superspace the hypermultiplet is 
described off shell by a Grassmann (G-)analytic superfield  
$q^+(x_A,\theta^+,\bar\theta^+,u^\pm)$. In it the harmonic 
variables are defined as $SU(2)$ matrices, 
\begin{equation}\label{1'}
  u\in SU(2)\quad \Rightarrow\quad   u^-_i =(u^{+i})^*\;, 
\quad u^{+i}u^-_i = 1   
\end{equation}
where the $SU(2)$ indices $i=1,2$ are raised and lowered in the 
usual way. All harmonic functions are homogeneous under the action 
of a $U(1)$ group counting the charges (like $\pm$ for $u^\pm$), 
so that they effectively live on the coset $S^2\sim U(1)\backslash 
SU(2)$. Further, the Grassmann variables  
\begin{equation}\label{1''}
  \theta^{+\alpha} = u^+_i\theta^{i\alpha},\quad 
\bar\theta^{+\dot\alpha} = u^+_i\bar\theta^{i\dot\alpha}  
\end{equation}
are $SU(2)$-invariant $U(1)$ projections of the full superspace 
ones $\theta^i,\bar\theta^i$. Finally, $x^{\alpha\dot\alpha}_A = 
x^{\alpha\dot\alpha} -4i \theta^{(i\, \alpha} 
  \bar\theta^{j)\, \dot\alpha}u^+_iu^-_j$ together with 
$\theta^+,\bar\theta^+$ and $u^\pm$ form a basis in the G-analytic 
superspace closed under the full $N=2$ superconformal group. 

It should be stressed that the off-shell hypermultiplet superfield 
$q^+(x_A,\theta^+,\bar\theta^+,u^\pm)$ involves an infinite number 
of auxiliary fields coming from its expansion on $S^2$. On shell 
$q^+$ obeys the free field equation 
\begin{equation}\label{2}
  D^{++}q^+(x_A,\theta^+,\bar\theta^+,u^\pm) = 0
\end{equation}
where 
\begin{equation}\label{3}
  D^{++} = u^{+i}\partial/\partial u^{-i} -2i\theta^{+\alpha}
\bar\theta^{+\dot\alpha} 
\partial/\partial x_A^{\alpha\dot\alpha}
\end{equation}
is the harmonic derivative in G-analytic superspace. Expanding 
equation (\ref{2}) in both the Grassmann and harmonic variables, 
one can show that all the auxiliary fields are eliminated and the 
remaining physical scalars and spinors are put on shell. 

The above off-shell formulation of the hypermultiplet allows one 
to develop standard Feynman rules \cite{hsgr}. In this context one 
can argue that the correlator (\ref{1}) obeys equations of the 
Schwinger-Dyson type, e.g., at point $r$, 
\begin{equation}\label{4}
   D^{++}_r\langle 
p_1p_2p_3p_4\rangle = \mbox{contact terms}\;, \quad 
r=1,\ldots,4\;. 
\end{equation}
In our analysis of superconformal covariants we always make the 
assumption that the points are kept apart in order to avoid 
space-time singularities. Then eq. (\ref{4}) becomes the condition 
of exact H-analyticity: 
\begin{equation}\label{5}
   D^{++}_r\langle 
p_1p_2p_3p_4\rangle = 0\;, \quad r=1,\ldots,4 \quad \mbox{if point 
1 $\neq\ldots\neq$ point $4$}\;. 
\end{equation} 

As we have seen in section 2, the combination of superconformal 
covariance and H-analyticity puts strong restrictions on the 
allowed form of such correlators expressed in terms of the 
coordinates of G-analytic superspace 
$x_A,\theta^+,\bar\theta^+,u^\pm$. However, such arguments cannot 
predict the dependence on the gauge coupling constant. The latter 
can be very efficiently determined by using the reduction formula 
of ref. \cite{ken1} (for an explanation in the context of $N=2$ 
harmonic superspace see \cite{hssw}). This formula relates the 
derivative of the $4$-point correlator (\ref{1}) with respect to 
the (complex) coupling constant $\tau$ to a $4+1$-point correlator 
obtained by inserting the $N=2$ SYM action: 
\begin{equation}\label{6}
  {\partial\over\partial \tau}\langle 
p_1p_2p_3p_4\rangle \sim \int d^4x_{L0} d^4\theta_0\langle {\rm 
tr}\;W^2(0)(q^+)^{p_1}(1)\ldots (q^+)^{p_4}(4)\rangle\;. 
\end{equation}
Here $W(x_L,\theta^{i\alpha})$ is the field strength of $N=2$ SYM 
and $L_{N=2\; SYM}= -{1\over 4} {\rm tr}\;W^2$ is the 
corresponding Lagrangian. Note that unlike the matter superfields 
$q^+$ which are G-analytic and harmonic-dependent off shell, $W$ 
is chiral and harmonic-independent. The integral in the reduction 
formula (\ref{6}) goes over the chiral insertion point 0. As we 
shall see later on, the combination of chirality with 
G-analyticity, in addition to conformal supersymmetry and 
H-analyticity, turns out to be extremely restrictive. 

So, from now on we shall concentrate on the $4+1$-point 
correlators 
\begin{equation}\label{7}
  \langle 0 p_1p_2p_3p_4\rangle 
\end{equation}
which are chiral at point 0 and G-analytic at points $1,\ldots,4$, 
have the corresponding superconformal properties and are also 
H-analytic, 
\begin{equation}\label{8}
   D^{++}_r\langle 0 
p_1p_2p_3p_4\rangle = 0\;, \quad r=1,\ldots,4 \quad \mbox{if point 
0 $\neq\ldots\neq$ point $4$}\;. 
\end{equation}  
In addition, they carry a certain $R$-weight. Indeed, the 
expansion of the matter superfield $q^+ = \phi^i(x)u^+_i + \ldots$ 
starts with the physical doublet of scalars of the $N=2$ 
hypermultiplet which have no $R$-weight. On the contrary, the 
$N=2$ SYM field strength $W = \ldots + \theta\sigma^{\mu\nu}\theta 
F_{\mu\nu}(x)$ contains the YM field strength $F_{\mu\nu}$ 
($R$-weight 0) in a term with two left-handed $\theta$'s, so the 
$R$-weight of $W$ equals 2 and that of the Lagrangian equals 4. 
>From (\ref{6}) it is clear that this weight is compensated by that 
of the chiral superspace measure $d^4x_Ld^4\theta$, so the 
correlator in the left-hand side of eq. (\ref{6}) is weightless. 

The task now is to explicitly construct superconformal covariants 
of $R$-weight 4 out of the coordinates of chiral superspace 
$x_{L0},\theta^{i\alpha}_0$ at the insertion point 0 and of 
G-analytic harmonic superspace $x_{A r},\theta^{+\alpha}_r, 
\bar\theta^{+\dot\alpha}_r, u^\pm_{ri}$, $r=1,\ldots,4$ at the 
matter points. To this end we need to know the transformation 
properties of these coordinates under $Q$ and $S$ supersymmetry 
(parameters $\epsilon$ and $\eta$, correspondingly) 
\cite{fradkin}: 
\begin{eqnarray}
 \delta x^{\alpha\dot\alpha}_L &=& -4i\theta^{i\alpha}\bar\epsilon^{\dot\alpha}_i
 -4i\theta^{i\alpha}x^{\beta\dot\alpha}_L
  \eta_{\beta i}  \nonumber\\
  \delta \theta^{i\alpha} &=& \epsilon^{i\alpha} + x^{\alpha\dot\beta}_L
\bar\eta^i_{\dot\beta} + 4i \theta^{i\alpha}\theta^{j\beta} 
\eta_{j\beta}\ ;  \label{9}\\ 
 && \nonumber\\
 \delta x^{\alpha\dot\alpha}_A &=& -4iu^-_i(\epsilon^{i\alpha}
\bar\theta^{+\dot\alpha} + 
\theta^{+\alpha}\bar\epsilon^{i\dot\alpha}) + 4i 
(x^{\alpha\dot\beta}_A\bar\theta^{+\dot\alpha}\bar\eta^i_{\dot\beta} 
-x^{\beta\dot\alpha}_A\theta^{+\alpha}\eta^i_\beta)u^-_i \nonumber 
\\ 
  \delta\theta^{+\alpha} &=& u^+_i\epsilon^{i\alpha} + 
 x^{\alpha\dot\beta}_A\bar\eta^i_{\dot\beta}u^+_i
-2i (\theta^+)^2\eta^{i\alpha}u^-_i \nonumber\\     
 \delta\bar\theta^{+\dot\alpha} &=& u^+_i\bar\epsilon^{i\dot\alpha} 
- x^{\beta\dot\alpha}_A\eta^i_{\beta}u^{+}_i +2i 
(\bar\theta^+)^2\bar\eta^{i\dot\alpha}u^-_i  \nonumber\\    
 \delta u^+_i&=& 4i
(\theta^{+\alpha}\eta^j_\alpha 
+\bar\eta^j_{\dot\alpha}\bar\theta^{+\dot\alpha})u^+_j u^-_i 
\nonumber\\ 
 \delta u^-_i&=&0 \;. \label{10}
\end{eqnarray}

We remark that $Q$ supersymmetry acts as a simple shift of the 
Grassmann variables, whereas $S$ supersymmetry is non-linear. 
Nevertheless, part of the $S$ transformation is shift-like (the 
terms $x^{\alpha\dot\beta}_L \bar\eta^i_{\dot\beta}$ in (\ref{9}) 
and $x^{\alpha\dot\beta}_A\bar\eta^i_{\dot\beta}u^+_i$, 
$x^{\beta\dot\alpha}_A\eta^i_{\beta}u^{+}_i$ in (\ref{10})), as 
long as we are allowed to invert the $x$'s. This is possible due 
to our choice to keep away from any singularities in $x$-space. 
Thus, we can use the four left-handed parameters 
$\epsilon^{i\alpha}$ and $x^{\alpha\dot\beta} 
\bar\eta^i_{\dot\beta}$ to shift away the two left-handed spinors 
$\theta^{i\alpha}_0$ at the chiral point (we count the $U(1)$ 
projections of the $SU(2)$ doublet $i=1,2$) and two of the 
left-handed $\theta^{+\alpha}_r$. This means that our correlators 
effectively depend on two left-handed spinor coordinates. We can 
make this counting argument more explicit by forming combinations 
of the $\theta$'s which are invariant under $Q$ supersymmetry and 
under the shift-like part of $S$ supersymmetry. $Q$ supersymmetry 
obviously suggests to use the combinations 
\begin{equation}\label{11}
  \theta^{\alpha}_{0r} = \theta^{i\alpha}_0u^+_{ri} - \theta^{+\alpha}_r\;, 
\quad \delta_Q \theta^{\alpha}_{0r} = 0\;,  \quad r=1,\ldots,4\;. 
\end{equation}
Then we can form the following two cyclic combinations of three 
$\theta^{\alpha}_{0r}$: 
\begin{equation}\label{12}
 (\xi_{12r})_{\dot\alpha} = (12)\rho_{r\dot\alpha} + 
(2r)\rho_{1\dot\alpha} + (r1)\rho_{2\dot\alpha}\;, \quad r=3,4  
\end{equation}
where
\begin{equation}\label{12'}
  \rho_{r\dot\alpha} = x^{-2}_{0r}(x_{0r}\theta_{0r})_{\dot\alpha}
\end{equation}
and $x_{0r}\equiv x_{L0}-x_{Ar}$ are translation-invariant and 
$(rs)\equiv u^{+i}_r u^+_{si}$ are $SU(2)$-invariant combinations 
of the space-time and harmonic coordinates, correspondingly. It is 
now easy to check that $\xi_{12r}$ are completely shift-invariant, 
i.e., 
\begin{equation}\label{13}
 \delta_{Q+S}\xi_{12r} = O(\theta^2)\;.
\end{equation}
Here one makes use of the harmonic cyclic identity 
\begin{equation}\label{131}
  (rs)t_i + (st)r_i + (tr)s_i = 0\;.
\end{equation}

Let us now inspect the structure of the correlator (\ref{7}) more 
closely. As we noted earlier, it has $R$-weight 4. In superspace 
the only objects carrying $R$-weight are the odd coordinates, so 
the $\theta$ expansion of our correlator must start with the 
product of four left-handed ones. In other words, the correlator 
should be nilpotent \cite{ehw,hssw}. Moreover, superconformal 
covariance requires that the shift-like transformations above do 
not reduce this number, so we must use all the four components of 
the shift-invariant combinations (\ref{12}) (notice that they have 
$R$-weight 1, even though they are right-handed spinors). Thus, we 
can write down the leading term in the correlator in the following 
form: 
\begin{equation}\label{14}
   \langle 0 
p_1p_2p_3p_4\rangle  = \xi_{123}^2 \xi_{124}^2 F^{p_1-4\vert 
p_2-4\vert p_3-2\vert p_4-2}(x,u) + O(\theta^5\bar\theta)\;. 
\end{equation}
The coefficient function $F$ depends on the space-time and 
harmonic variables and carries $U(1)$ charges to match those of 
the correlator and of the nilpotent prefactor. In the present 
context we are not interested in the purely conformal properties 
of this function, although it is easy to see that it should depend 
on the conformally invariant cross-ratios of the $x$'s (times a 
certain prefactor which gives the nilpotent term in 
(\ref{14}) the required dilation weight). 

Before going on we remark that the rest of the expansion 
(\ref{14}) is completely determined by superconformal covariance. 
Indeed, in order to keep the required $R$-weight we have to expand 
in pairs $\theta\bar\theta$. The shift-like transformations above 
do not mix $\bar\theta$ with $\theta$. Further, using the same 
counting argument as before, we conclude that there exists no 
nilpotent invariant made out of the four $\bar\theta^+$'s 
(chirality at point 0 prevents us from employing $\bar\theta_0$). 
So, all the higher-order terms in (\ref{14}) are uniquely related 
to the leading one by superconformal transformations. 

Besides superconformal covariance, the second main requirement on 
the correlator is H-analyticity (\ref{8}). Since we have only 
written out the leading term in the $\theta$ expansion (\ref{14}), 
there is no need to take into account the nilpotent part of the 
harmonic derivative (\ref{3}) (it only contributes to the next 
level in the expansion). Then H-analyticity is reduced to a 
condition concerning the pure harmonic dependence. The general 
solution of the H-analyticity condition on a harmonic function of 
charge $p$ is 
\begin{equation}\label{15}
  D^{++}f^p(u^\pm)=0 \ \Rightarrow \  \left\{ 
  \begin{array}{l}
    f^p = 0 \ \ \mbox{if}\ \ p<0; \\
    f^p = u^+_{i_1}\ldots u^+_{i_p}f^{i_1\ldots i_p} \ \ \mbox{if}\ \ p\geq 
0. 
  \end{array}
 \right.
\end{equation}
In other words, the solution only exists in the case of a 
non-negative charge and is a polynomial of degree $p$ in the 
harmonics $u^+$. The coefficient $f^{i_1\ldots i_p}$ forms an 
irrep of $SU(2)$ of isospin $p/2$. In our case, if we impose  
H-analyticity at a given point, the coefficient $f^{i_1\ldots 
i_p}$ can only be made out of the harmonics at the remaining 
points, since no other object carries $SU(2)$ indices (we have 
locked $\theta^{i\alpha}_0$ away in the combinations 
$\theta_{0r}$). Consequently, an H-analytic harmonic $4$-point 
function has the general form 
\begin{equation}\label{16}
  (12)^{m_{12}}(13)^{m_{13}}(14)^{m_{14}}(23)^{m_{23}}
  (24)^{m_{24}}(34)^{m_{34}}\;, \quad m_{rs}\geq 0\;. 
\end{equation}
  
Now, in our case we have to take account of the nilpotent 
prefactor in (\ref{14}). It would be too strong to demand that the  
function $F$ be H-analytic with respect to all the variables. The 
point is that the prefactor contains an overall factor $(12)^2$ 
which can improve the behaviour of the function $F$. Indeed, the 
detailed expansion of the prefactor has the form,
after some fierzing,

\begin{eqnarray}
 {\xi_{123}^2\xi_{124}^2\over (12)^2} &=&
(34)^2\rho_1^2\rho_2^2 
+2(23)(31)\rho_4^2(\rho_1\rho_2)
+{4\over 3}\bigl[(23)(14)+(13)(24)\bigr]
(\rho_1\rho_2)(\rho_3\rho_4)
\nonumber\\&&
+\,\,\mbox{all permutations}
\label{17} 
\end{eqnarray}
where $\rho_r$ were defined in (\ref{12'}). Then we can rewrite 
the expansion (\ref{14}) as follows: 
\begin{equation}\label{18}
   \langle 0 
p_1p_2p_3p_4\rangle  = {\xi_{123}^2\xi_{124}^2\over (12)^2} {\cal 
F}^{p_1-2\vert p_2-2\vert p_3-2\vert p_4-2}(x,u) + 
O(\theta^5\bar\theta)\;. 
\end{equation}
Since the expression (\ref{17}) has no harmonic zeros, it cannot 
suppress any singularities in the new coefficient function ${\cal 
F}$. So, we must require that ${\cal F}$ be H-analytic on its own, 
\begin{equation}\label{19}
  D^{++}_r {\cal 
F}^{p_1-2\vert p_2-2\vert p_3-2\vert p_4-2} = 0\;, \quad 
r=1,\ldots,4\;. 
\end{equation}
We have already seen that the general solution of such a 
condition, if it exists, is of the form (\ref{16}) where the total 
charges at each point should match those of ${\cal F}$. Thus we 
obtain the set of equations 
\begin{equation}\label{20}
  p_r-2 = \sum_{s\neq r}m_{rs}\;, \quad r=1,\ldots,4
\end{equation}
with $m_{rs}=m_{sr}$.  

The reason why eqs. (\ref{20}) do not always have a solution is 
that we are looking for $m_{rs}$ which are non-negative integers. 
This puts constraints on the allowed values of the correlator's 
charges $p_1,\ldots,p_4$. An obvious restriction is that the sum 
of all charges must be even. Further, it is easy to see that 
charges such that, e.g., 
\begin{equation}\label{21}
 p_4 >  p_1 + p_2 + p_3 - 4 
\end{equation}
are ruled out. A special case are the ``extremal" correlators of 
references \cite{dfmmrext,biakov} for which, e.g., 
\begin{equation}\label{22}
  p_4 = p_1 + p_2 + p_3\;. 
\end{equation}
We also see that no solution exists in the ``subextremal" case 
\begin{equation}\label{23}
   p_4 = p_1 + p_2 + p_3 - 2\;. 
\end{equation} 
 
Finally, we recall that the $4+1$-point correlators of the type 
considered here are uniquely determined by the leading term in 
their $\theta$ expansion. Therefore the constraints we have found 
apply to the entire superfield correlation function $\langle 0 
p_1p_2p_3p_4\rangle$, and, by means of the reduction formula 
(\ref{6}), to the $N=2$ matter correlators $\langle 
p_1p_2p_3p_4\rangle$. The conclusion is that neither the extremal 
nor the subextremal four-point correlators receive any quantum 
corrections beyond tree level. It should be mentioned that the 
method explained in section 2 leads to a slightly weaker condition 
on the coefficient functions of the subextremal correlators. It 
takes the form of a second-order PDE in the space-time variables 
which requires boundary conditions in order to fix the solution. 
In this context one might speculate that the reduction formula 
automatically takes account of some additional field-theory input 
which is harder to formulate in an approach based on symmetries 
alone.

To summarise, we have seen that the  simple expressions for 
extremal correlators which were derived in AdS supergravity in 
\cite{dfmmrext} can be easily understood on the field theory side 
as a consequence of analyticity in the harmonic superspace 
formalism. Furthermore, non-perturbative non-renormalisation 
theorems can be proven subject to the assumption that there are no 
unsuspected contact terms of the type which could interfere with 
the application of the reduction formula \cite{petske}. 
Strong arguments in 
favour of this assumption have been given in \cite{hssw}, 
including an explicit verification of the reduction formula at two 
loops in the symmetric case $p_1=\ldots = p_4 =2$. 

We would also like to emphasise that these results hold for 
analytic correlators in any $N=2$ SCFT and not just $N=4$, and for 
any choice of the gauge group. We further believe that these 
results can be extended to extremal correlators with an arbitrary 
number of points, although a complete proof would require a more 
detailed study of nilpotent covariants. This topic is under 
investigation.

\vspace{20pt} {\bf Acknowledgements:} CS would like to thank the 
Institute for Advanced Study, Princeton, for hospitality during 
the final stage of this project. ES and PW profited from 
stimulating discussions with M. Bianchi and E. D'Hoker. This work 
was supported in part by the British-French scientific programme 
Alliance (project 98074), by the EU network on Integrability, 
non-perturbative effects and symmetry in quantum field theory 
(FMRX-CT96-0012) and by the grant INTAS-96-0308.

\end{document}